\documentclass[
  aps,
  prd,superscriptaddress,
  nofootinbib,twocolumn,
  longbibliography
]{revtex4-2}

\usepackage{graphicx}
\usepackage{dcolumn}
\usepackage{bm}
\usepackage{mhchem}
\usepackage{xcolor}
\usepackage{hyperref}
\usepackage[english]{babel}
\usepackage[autostyle, english = american]{csquotes}
\MakeOuterQuote{"}

\begin{document}

\preprint{APS/123-QED}

\title{On the dynamical accessibility of twin stars}%

\author{Mahdi Naseri}%
 \email{mahdinaseri@arizona.edu}
\affiliation{Department of Astronomy, University of Arizona, Tucson, AZ, USA}

\author{Vasileios Paschalidis}%
 \email{vpaschal@arizona.edu }
\affiliation{Department of Astronomy, University of Arizona, Tucson, AZ, USA}
\affiliation{Department of Physics, University of Arizona, Tucson, AZ, USA}

\date{\today}

\begin{abstract}
A sufficiently strong hadron-to-quark first-order phase transition can give rise to a third family of stable compact stars that are commonly referred to as hybrid hadron-quark stars. Stable hybrid stars that have the same gravitational mass as neutron stars are referred to as twin stars. Although equilibrium twin stars may exist, whether they can be dynamically formed remains an open question. We investigate this problem by examining the gravitational binding energy of competing equilibrium configurations at fixed baryonic rest mass and by performing general relativistic hydrodynamical simulations of several possible transition channels. While twin stars are more gravitationally bound than neutron stars with the same rest mass, this energetic preference alone does not determine the dynamical outcome. Compression and shocks during the evolution generate thermal pressure that can prevent the system from settling on the cold twin star branch. Our simulations show that sufficiently rapid cooling can remove this thermal support and enable twin star formation, whereas slower or no cooling generally favors a neutron star remnant.  Accessing the twin star branch through the formation channels considered here requires cooling on a timescale comparable to or shorter than the stellar dynamical timescale. Since realistic cooling mechanisms operate on much longer timescales, our results suggest that in channels that conserve the total rest-mass neutron stars may be dynamically favored even when a more gravitationally bound twin star configuration exists with the same rest mass. Our results demonstrate a point of principle, at least for equations of state where the quark deconfinement density does not change appreciably for temperatures up to $10-20\,\textrm{MeV}.$
\end{abstract}

\maketitle


\section{\label{intro}Introduction}
Neutron stars (NSs) provide a unique laboratory for exploring the behavior of neutron-rich matter at densities exceeding the nuclear saturation density, $\rho_s\approx2.7\times 10^{14}$ g cm$^{-3}$. Under these extreme conditions, the properties of cold dense matter are governed by quantum chromodynamics (QCD), while the equation of state (EOS) remains poorly constrained at densities of $\sim 2-40\,\rho_s$ (see~\cite{Ozel:2016oaf,Lattimer:2021emm, Chatziioannou:2024jsr} for a review). According to QCD in sufficiently dense  or hot hadronic matter quark deconfinement can take place~\cite{Shuryak:1980tp,McLerran:1986zb}. An intriguing possibility is that a first-order hadron-quark phase transition could take place in the cores of NSs. When the surface tension between the two phases is sufficiently strong, a distinct class of stable compact objects can arise, which are known as hybrid hadron-quark stars, or hybrid stars (HSs)~\cite{Collins:1974ky, Heiselberg:1999mq,PhysRevGerlach,
Kampfer:1981yr, Glendenning:1998ag, Schertler:2000xq,
Ayvazyan:2013cva, Zacchi:2015oma, Bejger:2016emu,
Kaltenborn:2017hus, Alford:2017qgh, Blaschke:2018mqw,
AlvarezCastillo:2018pve, Blaschke:2020qqj,Orsaria:2013hna,Most:2018eaw,Ranea-Sandoval:2019miz,Otto:2019zjy, Tan:2021ahl,Pfaff:2021kse}. 

HSs consist of a quark matter core surrounded by a hadronic envelope, and on a mass--radius plot occupy a third stable branch, which is why they are also commonly referred to as the ``third family'' of compact objects [in addition to white dwarfs (WDs) and ordinary NSs].
The stable NS branch and the stable third family branch are separated by a branch of unstable configurations, analogous to the unstable WD branch separating stable WDs from stable NSs (see e.g.~Fig.~1 of Ref.~\cite{Naseri:2024rby} for a detailed description). A consequence of such a strong first-order phase transition is the possible existence of twin stars (TSs): pairs of stable NS and third family configurations with the same gravitational mass but different radii, with the TS being the more compact configuration~\cite{Zacchi:2016tjw,
AlvarezCastillo:2016oln, Christian:2017jni,Paschalidis:2017qmb, Montana:2018bkb,
Sieniawska:2018zzj, Sharifi:2021ead, Sen:2022lig,Tsaloukidis:2022rus}. 

The existence of a third family and TSs would have profound implications for nuclear physics and astrophysics and has motivated extensive recent theoretical work and searches for observational signatures capable of distinguishing HSs from purely hadronic NSs, see, e.g.,~\cite{Blaschke:2013ana,Bauswein:2018bma,Gieg:2019yzq,Weih:2019xvw,Bozzola:2019tit,Annala:2021gom,Prakash:2021wpz,Tan:2021nat,Haque:2022dsc,Blacker:2023afl,Lin:2023cbo,Guo:2023som,Pal:2023quk,Pradhan:2023zmg,Zhang:2024npg,Chan:2024nns,Fujimoto:2024ymt,Gartlein:2024cbj,Huang:2024xff,Jimenez:2024hib,Chatterjee:2025zrh,Zenati:2025wrc}. Measurements of NS properties have also been widely used to constrain the dense matter EOS and probe the possible presence of a hadron-to-quark phase transition~\cite{Chatziioannou:2019yko,Blacker:2020nlq,Xie:2020rwg,Miao:2020yjk,Tang:2020koz,Li:2020wbw,Ayriyan:2021prr,Takatsy:2023xzf,Christian:2023hez,Blacker:2024tet, Li:2024sft,Grundler:2025mcz,Kourmpetis:2025zjz,Li:2025tku,Haque:2026dre,Dong:2026kpb}. Recent constraints on the dense matter EOS from nuclear experiments and multimessenger observations, including measurements of NS masses and radii and tidal deformability constraints from gravitational wave observations, have been interpreted by several studies as being consistent with, or even favoring, a hadron-to-quark phase transition in NS cores~\cite{Paschalidis:2017qmb,Annala:2019puf,Annala:2021gom,Li:2022ivt,Essick:2023fso,Laskos-Patkos:2023tlr,Yamamoto:2023osc,Li:2024lmd,Laskos-Patkos:2024fdp,Mariani:2024gqi,Pal:2025skz}. However, other analyses find no preference for such a phase transition~\cite{Somasundaram:2021clp,Brandes:2023hma}. Therefore, whether HSs exist in nature remains an open question, and it is interesting to explore how these configurations, especially TSs, might form~\cite{Naseri:2024rby}.

Although stable TS solutions exist, equilibrium calculations alone do not determine which configurations are dynamically realized in nature. In particular, the existence of both stable branches of NSs and TSs raises the question of whether stellar evolution and gravitational collapse can populate both branches. This question has been investigated through several dynamical scenarios. General relativistic simulations have shown that configurations initially placed on the unstable branch separating stable NSs and TSs migrate toward the stable NS branch, rather than the stable TS branch, under a wide range of perturbations~\cite{Espino:2021adh}. Similarly, the gravitational collapse of stellar cores modeled as unstable WDs with rest masses in the TS range was found to produce stable NSs rather than HSs~\cite{Naseri:2024rby}. More recently, simulations that impose a cold, barotropic hybrid hadron-quark EOS probed the nonlinear response of stable NS and TS configurations to radial velocity perturbations~\cite{Haque:2026ero}. In that case, the two stable branches were found to respond differently to perturbations, with a critical perturbation required to induce migration depending on the stellar mass and the initial branch.

Despite considering different initial configurations and dynamical processes, Refs.~\cite{Espino:2021adh,Naseri:2024rby} raised the question: what determines the preferred outcome when multiple stable equilibrium configurations are available? In particular, the systematic migration toward the NS branch found in Refs.~\cite{Espino:2021adh,Naseri:2024rby}, together with the asymmetric response of stable NSs and TSs to perturbations reported in Ref.~\cite{Haque:2026ero}, suggests that a binding energy argument alone does not provide a complete description of the preferred configuration. In this work, we present simulations in full general relativity (GR) demonstrating that while these different results can be understood through the gravitational
binding energy of the competing equilibrium configurations, binding energy arguments are only part of the puzzle. We  demonstrate that the dynamical accessibility of the more gravitationally bound configurations is controlled by the evolution of the thermal energy generated during the evolution.

The remainder of this paper is organized as follows. In Sec.~\ref{energy}, we
review the gravitational binding energy of compact stars and develop our binding energy interpretation of the dynamical results discussed above. In Sec.~\ref{simulations}, we describe our numerical experiments and present the results used to test this interpretation. Finally, we conclude and summarize our unified picture in Sec.~\ref{conclusions}. Unless stated otherwise, throughout this work we adopt geometrized units where $c=G=1$, with $c$ denoting the speed of light in vacuum and $G$ representing the gravitational constant. In these units, one solar mass in length corresponds to
$GM_\odot/c^2 \approx 1.477~{\rm km}$.

\section{\label{energy}Binding energy analysis} 
\subsection{Binding energy definitions}

Three quantities are relevant in defining a binding energy in the asymptotically flat spacetime of compact stars: the gravitational or Arnowit-Deser-Misner (ADM) mass $M$, the rest mass $M_0$, and the proper mass $M_p$. 

The ADM mass for asymptotically flat systems is given by the following surface integral~\cite{PhysRevD.10.2345,Wald:1984rg}
\begin{equation} \label{ADM_mass}
M=\frac{1}{16\pi}\lim_{r \to \infty}\oint_{S_r}[\gamma^{jk}(\partial_k\gamma_{ij}-\partial_i\gamma_{jk})]\,dS^i \; .
\end{equation}
Here, $\gamma_{ij}$ is the spatial metric of 3D spatial hypersurfaces. The total rest mass of a star can be written as
\begin{equation} \label{rest_mass}
M_0=\int \rho_0 W\sqrt{\gamma} d^3x \; ,
\end{equation}
where $\rho_0$ is the rest mass density, $\gamma$ is the determinant of the spatial metric, and $W=\alpha u^0$ is the Lorentz factor measured by an observer  normal to $t=\rm const.$ hypersurfaces. Here, $u^\mu$ is the fluid four-velocity and $\alpha$ is the lapse function. In addition to these two, there is also the proper mass defined as~\cite{1959ApJ...130..884C}
\begin{equation} \label{proper_mass}
M_p=\int \rho W\sqrt{\gamma} d^3x \; ,
\end{equation}
where $\rho=\rho_0(1+e)$ is the total energy density and includes both the rest-mass and internal energy density ($e$ denotes the specific internal energy). Thus, $M_0$ accounts only for the baryonic rest-mass contribution, whereas $M_p$ additionally includes the internal energy of the fluid.  

The difference between the rest mass and the ADM mass defines the total binding energy~\cite{Prakash:1996xs,Lattimer:2000nx,Lattimer:2006xb},
\begin{equation} \label{binding_energy}
E_t=M-M_0 \; .
\end{equation}
The magnitude $|E_t|$ represents the net energy released in assembling the bound star from its constituent particles at infinity. Throughout this work we adopt the convention that the binding energy is negative for bound configurations. 

Similarly, the difference between the proper mass and rest mass defines the total internal energy,
\begin{equation} \label{nuclear_binding_energy}
U=M_p-M_0=\int \rho_0e W\sqrt{\gamma} d^3x \; .
\end{equation}

Finally, the gravitational binding (or potential) energy is defined as the difference between the ADM mass and proper mass~\cite{Paschalidis:2016vmz}, 
\begin{equation} \label{gravitational_binding_energy}
E_g=M-M_p=M-M_0-U \; ,
\end{equation}
and therefore,
\begin{equation} \label{Eb_and_E_t}
E_t=E_g+U \; .
\end{equation}

\begin{figure*}[t]
\includegraphics[width=\linewidth]{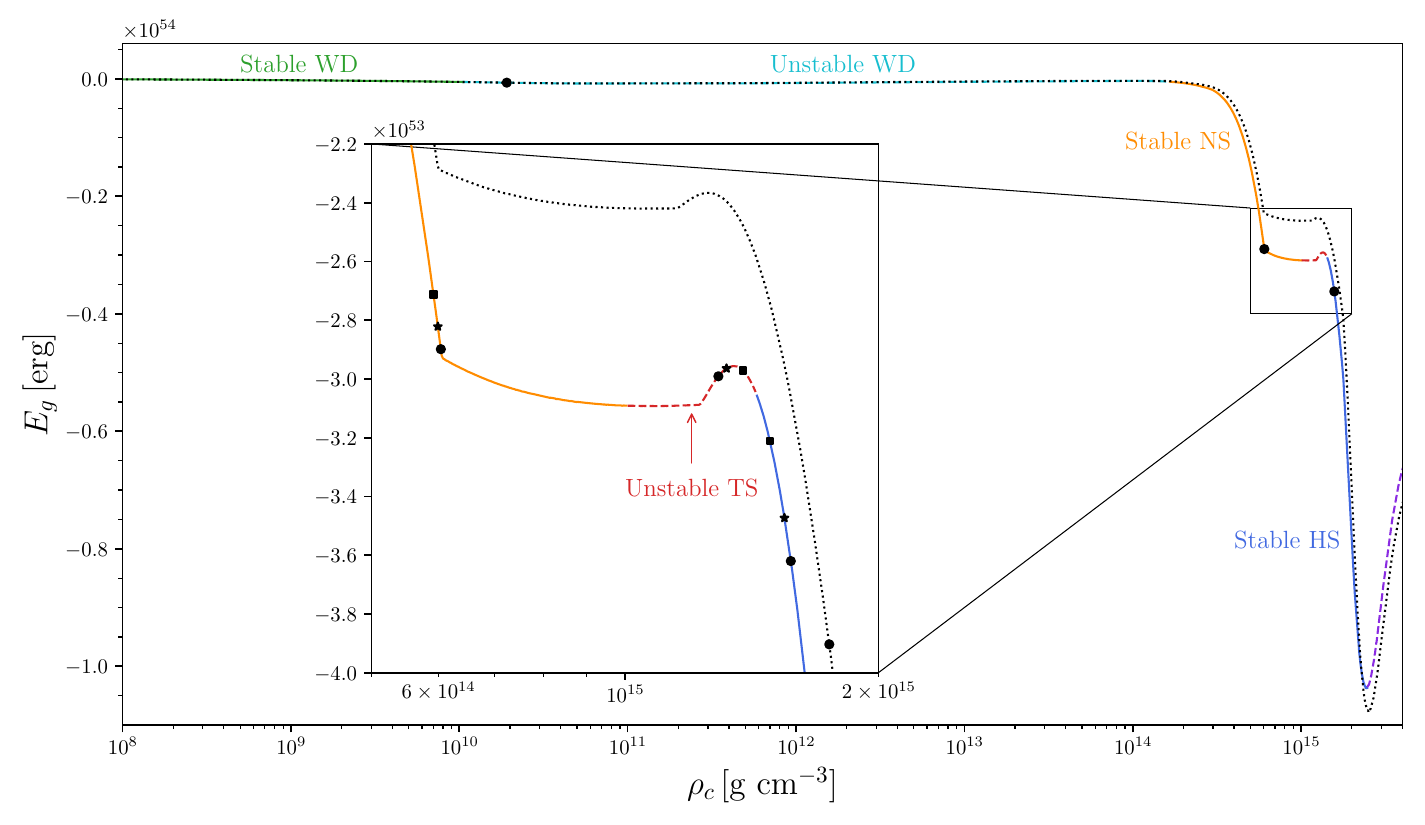}
\caption{\label{E_g} Gravitational binding of TOV configurations for the adopted hybrid hadron-quark EOS for a range of central densities covering all compact stars. Solid segments of the plot represent stable configurations, while the dashed lines represent the unstable solutions. Different branches of the curve are labeled by the corresponding color. The second curve, shown by dotted line, displays ``EOS II'' in~\cite{Naseri:2024rby}, which has no TS with $M_0=1.44\, M_\odot$. The points shown with a circle, asterisk, and square stand for stars with $M_0=1.44\, M_\odot$, $M_0=1.42\, M_\odot$, and $M_0=1.39\, M_\odot$, respectively.}
\end{figure*}

The last equation shows that the total binding energy contains two physically distinct contributions: the gravitational binding energy, $E_g$, associated with the self-gravity of the star, and the internal energy, $U$, determined by the microscopic state of the stellar matter. Rotational kinetic energy also contributes to the definition of binding energy, but in this work, the focus is on configurations with no rotation. 
$E_t$ and $E_g$ contain different physical information~\cite{2011MNRAS.413L..47B}; while $E_t$ represents the net binding of the star relative to its constituent baryons at infinity, $E_g$ isolates the gravitational contribution after accounting for
the local internal energy. This distinction becomes particularly important when comparing the binding energy of equilibrium configurations at fixed rest mass. From Eq.~\eqref{binding_energy}, it becomes clear that at fixed $M_0$, larger $E_t$ implies larger ADM mass, and vice versa. The same correspondence does not necessarily hold for $E_g$, because competing configurations can have different internal energies. As we demonstrate below, $E_g$ reveals an underlying energetic hierarchy among competing equilibrium configurations, where a TS is more bound than the corresponding NS. However, during a dynamical evolution, compression and shocks can generate additional internal energy and thermal pressure, so the star need not remain on the cold equilibrium sequence. Consequently, whether a more gravitationally bound configuration is dynamically accessible depends not only on the equilibrium energetic hierarchy, but also on the generation and removal of thermal energy during the evolution. The binding energy analysis is therefore useful as an initial guide, and we turn our attention to this next.

\subsection{Gravitational binding energy along TOV sequences}
For a given EOS, the Tolman--Oppenheimer--Volkoff (TOV) equations~\cite{Shapiro:1983du} determine a sequence of spherically symmetric hydrostatic equilibrium configurations parametrized by the central density $\rho_c$. The resulting $M$--$\rho_c$ relation can be used to identify changes in radial thermodynamic and dynamic stability through the turning-point theorem~\cite{Shapiro:1983du,Schiffrin:2013zta}. For the stellar sequences considered here, the stable WD branch is
followed by an unstable branch and, at higher central densities, by the stable NS branch, while at even higher central densities, the third family arises. Separating the stable NS branch from the stable HS branch, there is a branch of unstable stars. Hereafter, we refer to configurations on the unstable branch separating the stable NSs and stable HSs as unstable HSs, and 
we refer to unstable HSs with the same gravitational mass as stable NSs as unstable TSs.

For direct comparison with the dynamical studies discussed above, throughout this work we adopt the cold EOS labeled ``EOS I'' in Ref.~\cite{Naseri:2024rby}, which was also employed in Ref.~\cite{Haque:2026ero}. Once the equilibrium sequence is constructed, the quantities $M$, $M_0$, and $M_p$, and consequently $E_g$, can be evaluated for each configuration.

Fig.~\ref{E_g} shows the gravitational binding energy along a sequence of cold stars solving the TOV equations. To compare competing equilibrium configurations with the same baryonic mass, we select three fixed rest masses within the TS mass range: $M_0=1.44\,M_\odot$, $1.42\,M_\odot$, and $1.39\,M_\odot$, representing cases with high, intermediate, and low masses in the TS range considered for this particular EOS, respectively. For each value of $M_0$, multiple equilibrium configurations exist at different central densities which correspond to different families.

A clear ordering in gravitational binding energy emerges. At fixed $M_0$, the unstable WD is the least gravitationally bound configuration, followed by the stable NS, and then the unstable TS, while the stable TS is the most gravitationally bound. In our sign convention, this corresponds to
\[
E_g^{\rm stable\;TS}<E_g^{\rm unstable\;TS}<E_g^{\rm stable\;NS}<E_g^{\rm unstable\;WD}.
\]
The corresponding values are reported in Table~\ref{tab:binding_masses}, which suggests the same ordering for all three rest masses considered. Thus, among the competing configurations at fixed $M_0$, increasing central density is accompanied by increasingly negative $E_g$, with
the stable TS representing the most gravitationally bound equilibrium configuration. This behavior is consistent with the increasing compactness of the corresponding configurations: at a given rest mass, the more compact solutions generally possess a larger magnitude of gravitational binding energy. 

\begin{figure}[h]
\includegraphics[width=\linewidth]{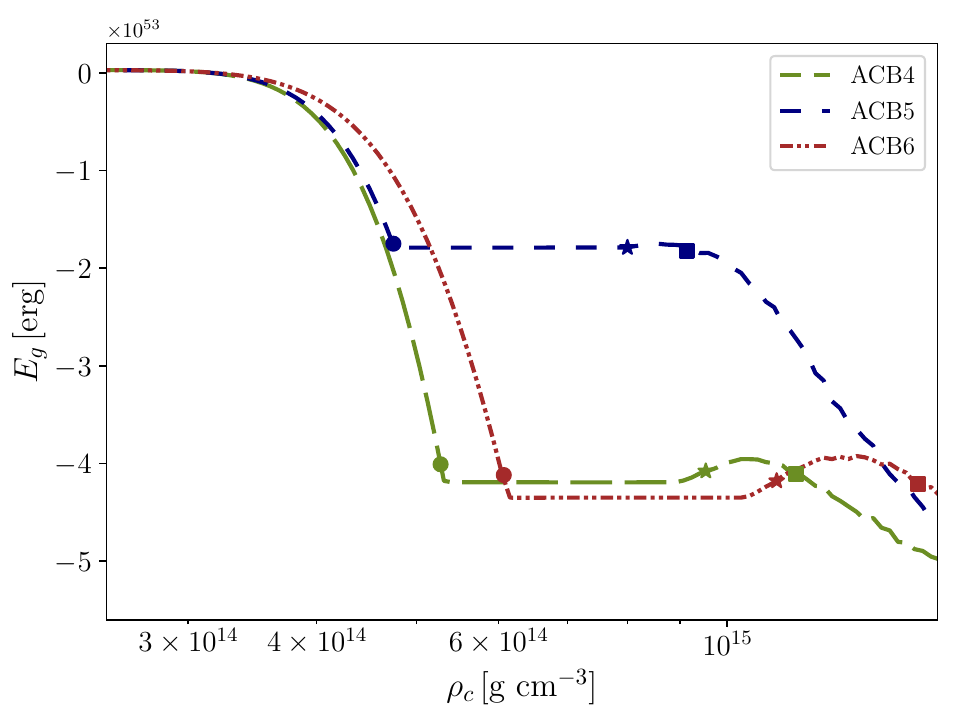}
\caption{\label{E_g_general} Gravitational binding energy of a TOV sequence with various  EOSs that exhibit a strong first-order hadron-to-quark phase transition. The EOSs shown on this plot are taken from Ref.~\cite{Paschalidis:2017qmb} using the same labeling. At crust densities, they are smoothly extended using the low-density pieces of ``EOS I'' from Ref.~\cite{Naseri:2024rby}. The curves are only shown for the density range of TSs. For each EOS, three cold configurations with a fixed rest mass are illustrated; they are a stable NS (circle), an unstable TS (asterisk), and a stable TS (square). For all the EOSs shown, the overall behavior of the plot is similar to Fig.~\ref{E_g}, and the same $E_g$ ordering is obtained.}  
\end{figure}

To examine whether this behavior is specific to our EOS, in Fig.~\ref{E_g_general} we repeat the calculation for several additional EOSs that give rise to a third family branch. The same qualitative ordering is obtained, suggesting that the behavior identified above is not peculiar to the EOS used in our dynamical simulations. 

The total binding energy exhibits a qualitatively different behavior. Table~\ref{tab:binding_masses} also lists the total binding energy of the stable NS, unstable TS, and stable TS configurations at the three selected values of $M_0$. Unlike $E_g$, which
decreases monotonically across these configurations with increasing central density, $E_t$ is non-monotonic: the unstable TS lies at or near
a local maximum in $E_t$ between the two stable branches. 

The contrasting behaviors of $E_g$ and $E_t$ reflect the different physical information contained in the two quantities. The former reveals a monotonic hierarchy in the gravitational binding energy of the
competing configurations, whereas the latter describes their net binding energy after the internal energy contribution is included. Along a
cold equilibrium sequence, this internal energy is dictated by the EOS. However, during a dynamical transition, compression and shocks can generate additional internal thermal energy and thermal pressure, so the star need not remain on the cold equilibrium sequence. Consequently, the cold equilibrium sequence binding energy alone does not determine which
configuration is dynamically accessible or more favorable. In the following section, we conduct numerical experiments to investigate how the gravitational binding energy and thermal evolution together determine the outcome of transitions between the different branches.

\begin{table}[h]
\caption{\label{tab:binding_masses}
The magnitude of the total binding energy  $|E_t|$ and the gravitational binding energy $|E_g|$ of the stable NS, unstable TS, and stable TS configurations for selected
values of fixed rest mass $M_0$. All quantities are given in units of $M_\odot$.}
\begin{ruledtabular}
\begin{tabular}{l|cc|cc|cc}
& \multicolumn{2}{c|}{Stable NS}
& \multicolumn{2}{c|}{Unstable TS}
& \multicolumn{2}{c}{Stable TS} \\
$M_0$ & $|E_t|$ & $|E_g|$
      & $|E_t|$ & $|E_g|$
      & $|E_t|$ & $|E_g|$ \\
\colrule
1.44 & 0.1008 & 0.1621 & 0.1008 & 0.1673 & 0.1020 & 0.2025 \\
1.42 & 0.0978 & 0.1578 & 0.0975 & 0.1658 & 0.0982 & 0.1943 \\
1.39 & 0.0933 & 0.1517 & 0.0927 & 0.1662 & 0.0928 & 0.1796 \\
\end{tabular}
\end{ruledtabular}
\end{table}

\section{\label{simulations}Numerical Experiments} 

\subsection{Energetics of dynamical transitions}
The gravitational binding energy landscape discussed above provides a useful framework for interpreting the dynamical results reviewed in Sec.~\ref{intro}. At fixed rest mass, Fig.~\ref{E_g} shows that the gravitational binding energy becomes progressively more negative when moving from the unstable WD branch toward the stable NS, unstable TS, and stable TS branches. Therefore, among the equilibrium configurations available at a given rest mass, the stable TS represents the most gravitationally bound state. Nevertheless, previous dynamical simulations do not necessarily evolve toward this configuration. In particular, unstable TSs were found to migrate toward stable NSs~\cite{Espino:2021adh}, and the collapse of unstable WDs with rest masses in the TS range similarly resulted in stable NSs~\cite{Naseri:2024rby}. These results indicate that the existence of a more gravitationally bound equilibrium configuration does not by itself guarantee that it will be reached dynamically.

The situation becomes interesting for initially stable configurations. Ref.~\cite{Haque:2026ero} investigated stable NSs and TSs subjected to radial velocity perturbations and found that sufficiently strong perturbations can induce transitions between the two branches. However, the evolutions in Ref.~\cite{Haque:2026ero} were performed using the cold barotropic EOS throughout the evolution, thereby preventing the generation of heat. This corresponds to the limiting case in which any dynamically generated thermal contribution is removed instantaneously, so that the fluid remains on the cold EOS. However, this is not the case in nature. Compression and shocks can generate heat, and radiating the excess thermal energy away does not take place instantaneously. This raises the question of how the transition changes when dynamically generated thermal energy is retained in the fluid, and how the thermal evolution impacts the ultimate fate of the perturbed star.

Although the initial equilibrium configurations are constructed using a cold EOS, a dynamically evolving star does not, in general, remain on the zero-temperature equilibrium sequence. A velocity perturbation
injects bulk kinetic energy, while contraction changes the internal energy as the fluid is compressed. In  nonlinear
evolutions, shocks can additionally convert bulk kinetic energy into thermal energy, generating a nonzero thermal component and an associated thermal pressure. Consequently, the internal energy $U$ is
no longer restricted to its value along the cold equilibrium sequence. Through Eq.~\eqref{Eb_and_E_t}, changes in the thermal
content alter the relation between the total and gravitational binding energies. Moreover, any bulk kinetic energy must be taken into account in the binding energy calculation. In the absence of an efficient cooling mechanism, the generated thermal energy remains largely trapped within the star and
can provide additional pressure support against further contraction. If cooling is present, part of this energy can instead be radiated away, allowing the configuration to continue contracting and potentially settle onto a more compact equilibrium branch.

These considerations motivate the central hypothesis of our work: the ordering in $E_g$ provides an underlying energetic hierarchy among the competing equilibrium configurations, while the generation and removal of thermal energy determine which of these configurations are dynamically accessible. To test our hypothesis we  perform controlled numerical experiments in
which we allow heat to be generated, but we account for cooling, with varying cooling timescale. This allows us to determine whether transitions toward the more gravitationally bound branches require the dynamically generated thermal energy to be removed, and to identify the conditions under which such transitions can occur.

\subsection{Computational setup}
Our numerical simulations are performed within the \texttt{Einstein Toolkit} infrastructure~\cite{Loffler:2011ay,EinsteinToolkit:2023_05}, using \texttt{Cactus}~\cite{Goodale2002a} and \texttt{Carpet}~\cite{Schnetter:2003rb} for adaptive mesh refinement, with simulations managed through \texttt{SimFactory}~\cite{sim}. The tabulated EOS and TOV equilibrium sequence are constructed independently using our own Python codes as in Ref.~\cite{Naseri:2024rby}. Simulation data are analyzed using the \texttt{kuibit} post-processing package~\cite{Bozzola:2021hus}.

The stellar initial data are generated with the \texttt{RNSID} thorn, which is based on the \texttt{RNS} code~\cite{1995ApJ444306S,Stergioulas:2003yp}. \texttt{RNSID} constructs isolated equilibrium stellar configurations assuming a zero-temperature EOS supplied in either tabulated or polytropic form. Since the piecewise polytropic EOS adopted in this work is not directly supported in the required form, we construct a highly sampled tabulated representation of the EOS and provide it to \texttt{RNSID}. Although \texttt{RNSID} allows for rotating configurations, all stars considered here are non-rotating. 

The spacetime is evolved using the Baumgarte--Shapiro--Shibata--Nakamura (BSSN) formulation~\cite{Baumgarte:1998te,Shibata:1995we}, as implemented in the \texttt{Lean} code~\cite{Sperhake:2006cy}. We employ the standard ``$1+\log$'' slicing condition and the "$\Gamma$-driver" shift condition~\cite{Bona:1994dr,Alcubierre:2002kk}. The hydrodynamic equations are evolved with \texttt{IllinoisGRMHD}~\cite{Etienne:2015cea,github_GRMHD}. 

During the evolution, we employ a hybrid EOS consisting of cold and thermal components,
\begin{equation}
P=P_{\rm cold}+P_{\rm th},
\label{cold_th}
\end{equation}
where $P_{\rm cold}$ is determined by the zero-temperature piecewise polytropic EOS and $P_{\rm th}$ accounts for thermal pressure generated during the evolution. The latter is modeled using a $\Gamma$-law equation,
\begin{equation}
P_{\rm th}=(\Gamma_{\rm th}-1)\rho_0e_{\rm th},
\end{equation}
with $e_{\rm th}=e-e_{\rm cold}$ and $\Gamma_{th}=2$. Here, $e_{\rm cold}$ is the zero-temperature specific internal energy determined by the cold EOS. This decomposition allows the fluid to depart from the cold EOS during the dynamical evolution and, in particular, captures the thermal energy and pressure generated by shocks in the phase transition region and compression.

To trigger transitions away from equilibrium, we apply controlled pressure and radial velocity perturbations to the initial stellar configurations. Pressure perturbations are introduced according to
\begin{equation}
P\rightarrow (1+\xi_p)P =P+\delta P\; ,
\label{pressure_pert}
\end{equation}
where $\xi_p
={\delta P}\slash{P}$ controls the perturbation amplitude. Negative values of $\xi_p$ correspond to pressure depletion and favor contraction, whereas positive values increase the initial pressure support. Radial uniform velocity perturbations are imposed as
\begin{equation}
\vec{v}\rightarrow \xi_v \, \hat{r},
\label{velocity_pert}
\end{equation}
where $\xi_v$ determines the amplitude and direction of the perturbation (we only use inward velocity perturbations in this work, which corresponds to $\xi_v <0$). This uniform velocity perturbation is consistent with the perturbations used in Ref.~\cite{Haque:2026ero}. 

To control the thermal energy generated during the evolution, we implement the effective cooling prescription of Ref.~\cite{Paschalidis:2011ez} in \texttt{IllinoisGRMHD}, as adopted in Ref.~\cite{Naseri:2024rby}. The local cooling emissivity is taken to be
\begin{equation}
\Lambda=\frac{\rho_0 e_{\rm th}}{\tau_c},
\label{Lambda}
\end{equation}
where $\tau_c$ is the cooling timescale. The corresponding source terms are added to the energy and momentum equations as described in Ref.~\cite{Paschalidis:2011ez}; we refer to
Ref.~\cite{Naseri:2024rby} for details of our implementation. The cooling timescale is set relative to the dynamical timescale of the initial configuration defined as $t_{dyn}\equiv 1/\sqrt{\rho_c(t=0)}$. In the absence of compression or expansion, this prescription leads to an exponential decay of the thermal component, $e_{\rm th}\propto \exp(-\tau /\tau_c)$, with $\tau$ the fluid proper time. Thus, cooling drives the fluid toward the zero-temperature EOS on a controllable timescale.

We employ the same grid structures and resolutions as in Ref.~\cite{Naseri:2024rby}. For simulations starting from an unstable WD, the initial grid consists of two refinement levels, with the finer
level resolving the stellar radius by 100 grid points. As the star contracts and its central density increases, six additional refinement
levels are activated sequentially, with each new level doubling the spatial resolution. When all refinement levels are active, the finest grid spacing is approximately $166\,{\rm m}$. The outer boundary is located at $1.7R_{\rm WD}\simeq498M_{\rm WD}$. For simulations starting from compact NS and TS configurations, we employ 7-8 refinement levels depending on the radius of the initial configuration, with the requirement that the finest level resolves the initial stellar radius by $\simeq100$ grid points. Further details of the grid setup and refinement strategy can be found in Ref.~\cite{Naseri:2024rby}.

\subsection{Results}

\subsubsection{Migration of unstable twin stars}
We first consider configurations on the unstable branch separating the stable NS and TS branches in Fig.~\ref{E_g}, corresponding to the class of initial data investigated in Ref.~\cite{Espino:2021adh}. We select three configurations with rest masses $M_0=1.44\,M_\odot$, $1.42\,M_\odot$, and $1.39\,M_\odot$, denoted by circles, asterisks, and squares, respectively, in Fig.~\ref{E_g}. These cases span the TS mass range of the adopted EOS. As discussed earlier, for each  rest mass the gravitational binding energy decreases from the stable NS to the unstable TS and finally to the stable TS. We therefore use these configurations to investigate why an unstable TS does not necessarily evolve toward the most gravitationally bound equilibrium configuration and how this behavior is affected by the removal of excess thermal energy.

\begin{figure}[t]
\includegraphics[width=\linewidth]{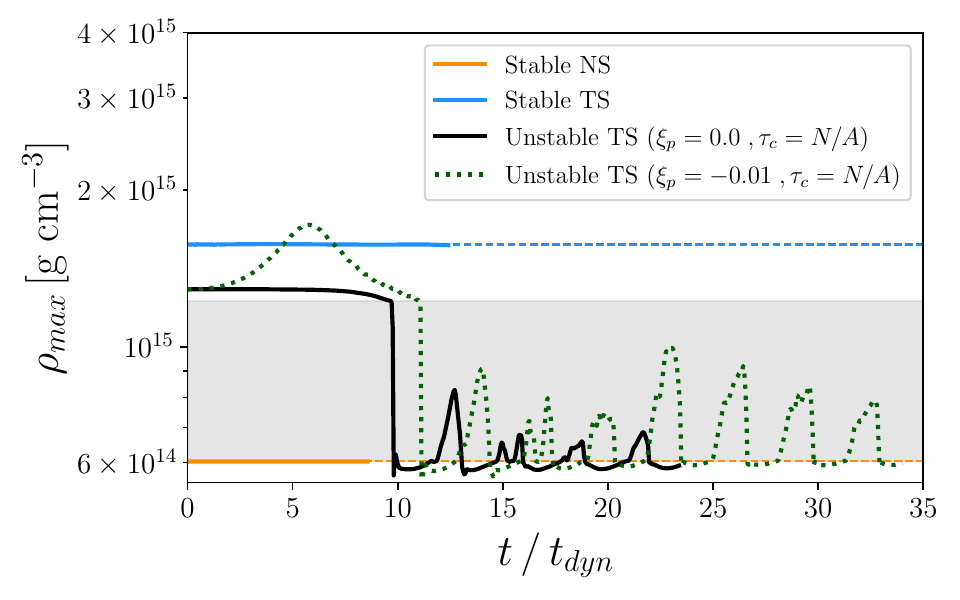}
\caption{\label{UTS_44} Maximum density evolution of the unstable TS with $M_0=1.44\,M_\odot$ with and without initial perturbation in pressure. In all the plots throughout, time is scaled by the dynamical timescale $t_{dyn}$ of the initial configuration, the shaded area denotes the phase transition range of densities, and $\tau_c = N/A$ means that the cooling prescription is not activated. In addition, orange and blue dashed lines represent the maximum density of the stable NS and TS with the same rest mass as the other curves, respectively. In this plot, the solid horizontal lines also show the actual simulation data for the two stable stars, which indicate that stability is maintained.} 
\end{figure}

Fig.~\ref{UTS_44} shows the evolution of the maximum rest-mass density for the $M_0=1.44\,M_\odot$ configuration in the absence of cooling. For reference, the central densities of the stable NS and stable TS with the same rest mass are indicated by the orange and blue horizontal lines, respectively, while the shaded region marks the phase transition density range. The unstable initial configuration lies between the two stable branches.

Because this configuration is dynamically unstable, even numerical perturbations at the grid level are sufficient to drive it away from equilibrium. In the unperturbed evolution, the maximum density decreases and the star migrates toward the corresponding stable NS. When a $1\%$ pressure depletion is imposed, the star initially contracts toward the TS branch. However, after reaching its maximum density, it re-expands, crosses the phase transition region, and subsequently oscillates around the stable NS configuration. Thus, in the absence of cooling, an initial motion toward higher density is not sufficient for the star to settle on the stable TS branch.

To assess the robustness of this result with resolution, we repeated the pressure-perturbed case (shown by dotted curve in Fig.~\ref{UTS_44}) at twice the spatial resolution, resolving the stellar radius by 200 grid points, and obtained the same result. The agreement between the two resolutions, particularly in the stable branch ultimately approached by the star, demonstrates that the qualitative outcome relevant to the present analysis is robust with resolution. Furthermore, the solid orange and blue curves display the evolution of maximum density for the stable NS and TS configurations with $M_0=1.44\,M_\odot$, respectively. These two cases exhibit small fluctuations ($\lesssim 0.01-0.1\%$) around the initial maximum density, showing that stable equilibria are preserved during the evolution.

\begin{figure}[t]
\includegraphics[width=\linewidth]{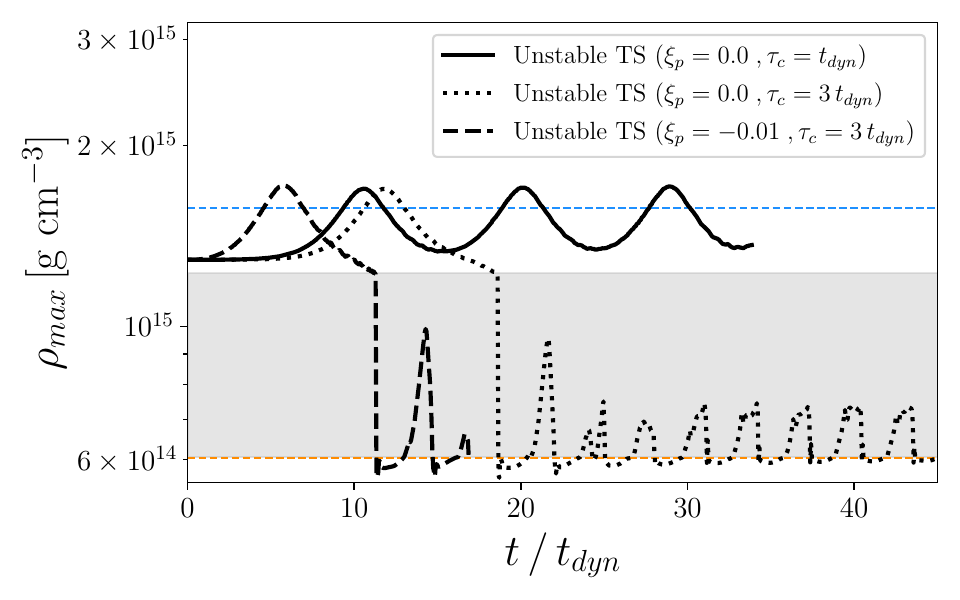}
\caption{\label{UTStoHS_44} The fate of unstable TSs with $M_0=1.44\,M_\odot$ having cooling activated in the evolution with different choices of cooling timescale and initial pressure perturbation. Orange and blue horizontal lines represent the maximum density of the corresponding stable NS and TS, respectively.} 
\end{figure}

The outcome changes when thermal energy is removed during the evolution. Fig.~\ref{UTStoHS_44} shows the same $M_0=1.44\,M_\odot$ unstable TS evolved with different cooling timescales. For rapid cooling, $\tau_c=t_{\rm dyn}$, the star remains on the high-density side of the phase transition region and ultimately oscillates around the stable TS configuration. In contrast, for the slower cooling timescale $\tau_c=3\,t_{\rm dyn}$, the star re-expands through the phase-transition region and approaches the stable NS branch. Adding an initial pressure perturbation only expedites the process but does not qualitatively change this outcome. These simulations therefore demonstrate that cooling can enable formation of a stable TS, but only when thermal energy is removed sufficiently rapidly during the initial contraction. 

\begin{figure}[h]
\includegraphics[width=\linewidth]{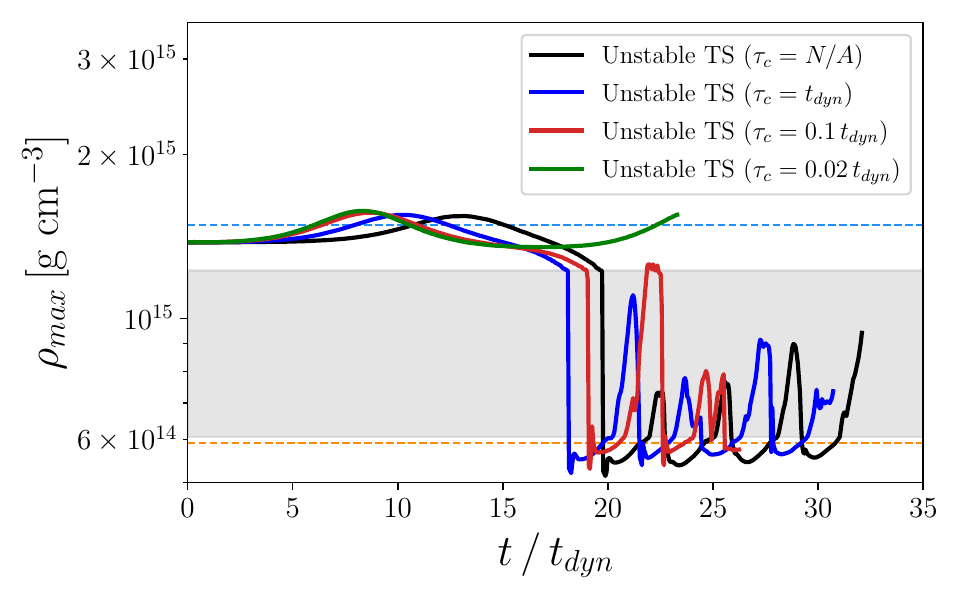}
\caption{\label{UTS_39} The evolution of an unstable TS with $M_0=1.39\,M_\odot$ with no pressure perturbation but with different choices for cooling timescale. The dashed orange and blue lines stand for the maximum density of stable NS and TS with $M_0=1.39\,M_\odot$, respectively.} 
\end{figure}

\begin{figure}[h]
\includegraphics[width=\linewidth]{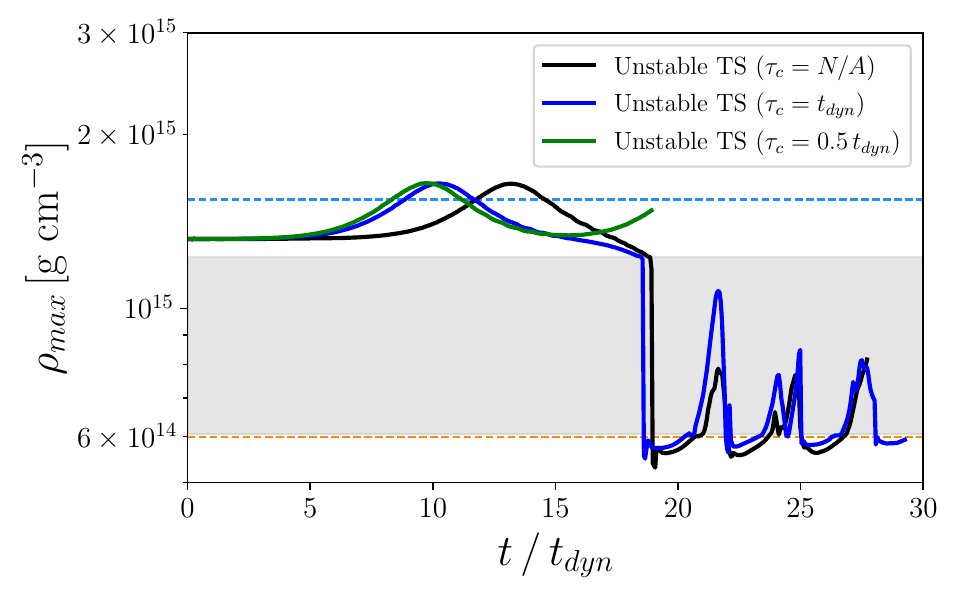}
\caption{\label{UTS_42} The evolution of an unstable TS with $M_0=1.42\,M_\odot$ without any pressure perturbation but with three different choices for cooling timescale. Similar to the previous plots, the maximum density of stable NS and TS with $M_0=1.42\,M_\odot$ are depicted by the dashed blue and orange lines, respectively.} 
\end{figure}

Figs.~\ref{UTS_39} and~\ref{UTS_42} repeat this experiment for unstable TSs with $M_0=1.39\,M_\odot$ and $1.42\, M_\odot$, respectively. The same qualitative behavior is observed; sufficiently rapid cooling allows the unstable configuration to settle on the stable TS branch, whereas slower cooling results in migration toward the stable NS. Importantly, the required cooling timescale for the configuration to remain on the TS branch depends strongly on which unstable TS configuration we evolve. Looking at  Table~\ref{tab:binding_masses} there is a correlation of this cooling timescale 
with the total binding energy and the mass of the unstable TSs, becoming progressively shorter for the lower mass or lower $|E_t|$ unstable TSs. For $M_0=1.39\,M_\odot$, even $\tau_c=0.1\,t_{\rm dyn}$ is insufficient to produce a stable TS. Thus, the relevant criterion is not the presence or absence of cooling, but whether thermal energy can be removed sufficiently rapidly relative to the dynamical evolution of the star. In other words, if heat is instantly removed (as is effectively the tacit assumption in~\cite{Haque:2026ero}), then the TS branch can be reached.

\begin{figure*}[t]
\includegraphics[width=\linewidth]{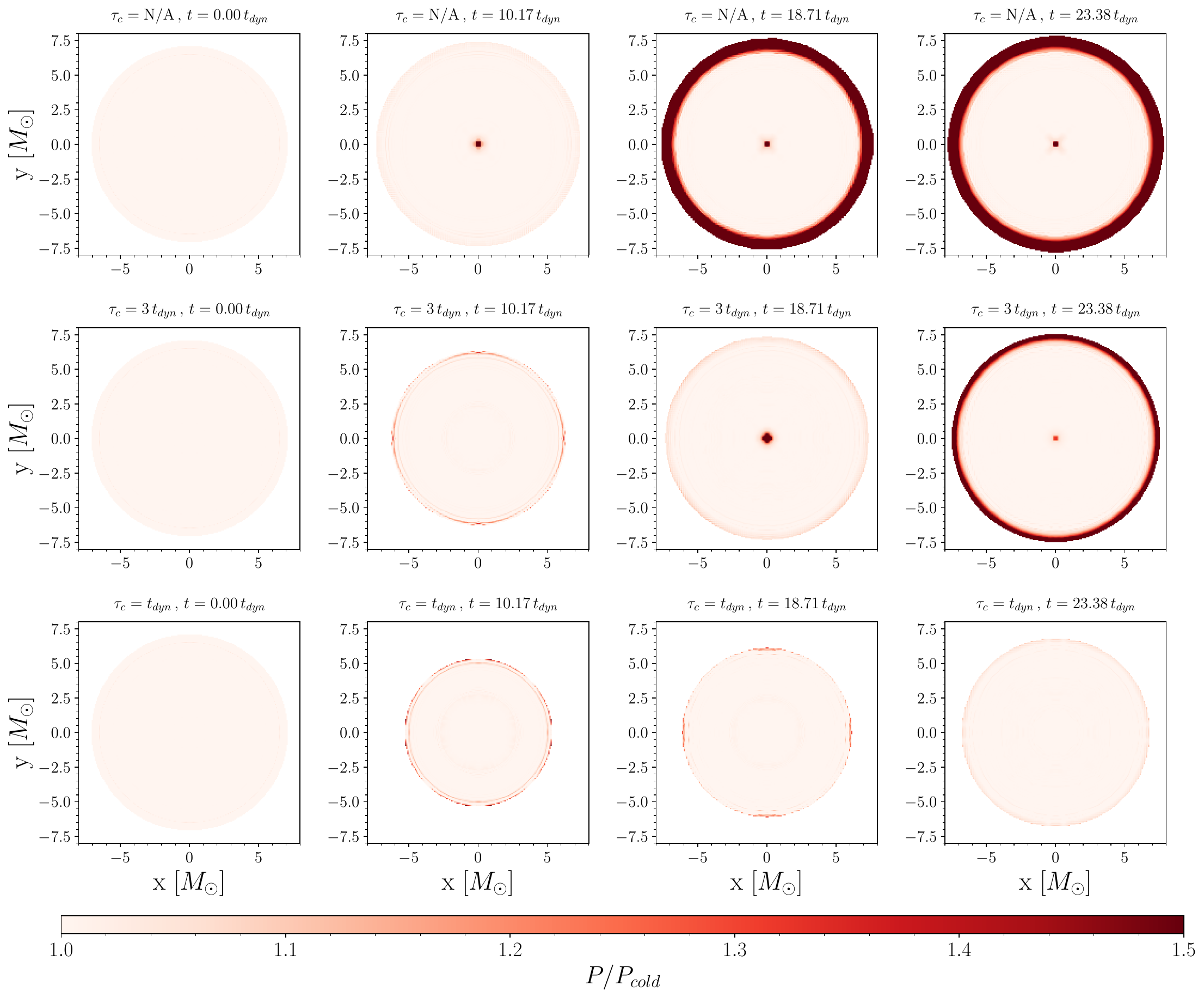}
\caption{\label{cooling} Snapshots of $P/P_{\rm cold}$ on the equatorial plane for the unstable TS with $M_0=1.44\,M_\odot$. It includes the solid black curve in Fig.~\ref{UTS_44}, i.e. no cooling (top plots), the dotted black curve in Fig.~\ref{UTStoHS_44},  i.e. slow cooling $\tau_c=3t_{\rm dyn}$ (middle plots), and the solid black curve in Fig.~\ref{UTStoHS_44}, i.e. rapid cooling $\tau_c=t_{\rm dyn}$ (bottom plots). All configurations are initially cold, $P/P_{\rm cold}=1$. Compression and shocks generate a thermal pressure component during the subsequent
evolution, while the cooling prescription removes this component on the prescribed timescale. The snapshots illustrate the different amounts of thermal support retained (especially near the quark core) as the star evolves toward the NS or TS branch. The plot shows $P/P_{\rm cold}$ for rest-mass densities above $10^{-3}\rho_{max}$ where the bulk of the stellar rest mass lies.} 
\end{figure*}

To understand the origin of this behavior, in Fig.~\ref{cooling} we examine the spatial distribution of $P/P_{\rm cold}$ for the $M_0=1.44\,M_\odot$ configuration. Since $P=P_{\rm cold}+P_{\rm th}$, departures of this ratio from unity provide a measure of the thermal pressure generated during the evolution. The three rows correspond to evolutions without cooling, with slow cooling, and with rapid cooling. Although all three configurations are initially cold, contraction and shocks generate a thermal component concentrated primarily in the stellar interior. Without cooling, this thermal energy remains in the fluid, while with cooling, it is progressively removed, with the amount retained during the contraction determined by the ratio of the cooling and dynamical timescales. 

These results also illustrate why the equilibrium ordering in $E_g$ does not by itself determine the dynamical outcome. Although the baryonic rest mass $M_0$ is conserved to a good approximation during the evolution, the star does not evolve along the sequence of cold equilibrium configurations. Contraction and shocks generate thermal energy and pressure, providing additional support against further contraction. This thermal support can
cause the star to re-expand before it can settle onto the stable TS branch, despite the stable TS being more gravitationally bound at the same rest mass.

Cooling changes this evolution by removing the thermal energy generated during contraction and shocks. If cooling operates rapidly enough, thermal support is reduced while the star remains on the high-density side of the phase transition region, allowing it to continue contracting and relax toward the cold stable TS. If cooling is too slow, thermal pressure causes the star to re-expand toward lower
densities before sufficient thermal energy has been removed, after which it relaxes toward the stable NS branch. The final outcome is therefore determined not by the equilibrium ordering of $E_g$ alone, but
by whether the thermal energy generated during the transition can be removed before it reverses the contraction. In other words, the outcome is not path independent.

\subsubsection{Transitions between stable neutron stars and twin stars}
We next consider transitions between initially stable NS and TS configurations. Unlike the unstable configurations studied above, these stars remain close to equilibrium in the absence of a sufficiently strong perturbation. A transition between the two stable branches therefore requires a finite perturbation capable of driving
the star through the intermediate unstable region.

Such transitions were recently studied in Ref.~\cite{Haque:2026ero}, where radial velocity perturbations were applied to stable NSs and TSs
to determine the critical perturbation required to induce migration from one branch to the other. The resulting thresholds were interpreted in terms of the total binding energy. 
However, an important feature of the simulations in
Ref.~\cite{Haque:2026ero} is that the EOS was restricted to be adiabatic. The dynamically generated thermal-pressure
response was therefore suppressed by construction. In an evolution with the hybrid EOS of Eq.~\eqref{cold_th}, part of the kinetic energy
introduced by the perturbation can instead be converted into internal energy and thermal pressure during compression and shock heating. The star can therefore depart from the cold equilibrium sequence,
altering its subsequent dynamical evolution.

We therefore repeat the velocity perturbation experiments of Ref.~\cite{Haque:2026ero}, but evolve the fluid using the full hybrid EOS of Eq.~\eqref{cold_th}, allowing a thermal component to develop dynamically. The migration from a stable TS to the stable NS branch following a sufficiently strong perturbation is naturally consistent
with the gravitational binding energy hierarchy: the perturbation injects energy into the initially more gravitationally bound TS and can drive it across the unstable region toward the less bound stable
NS branch.

The reverse transition, from a stable NS to a stable TS, is more subtle. At fixed rest mass, Fig.~\ref{E_g} shows that the stable TS is
more gravitationally bound than the corresponding stable NS. An inward velocity perturbation can initially compress the NS to densities above
the phase-transition region, but at the same time it injects kinetic energy into the system. During the subsequent evolution, part of this energy is converted into internal energy and thermal pressure. Thus, although the perturbation can initiate the transition toward higher density, it does not by itself allow the star to relax to the more
gravitationally bound cold TS configuration.

\begin{figure}[t]
\includegraphics[width=\linewidth]{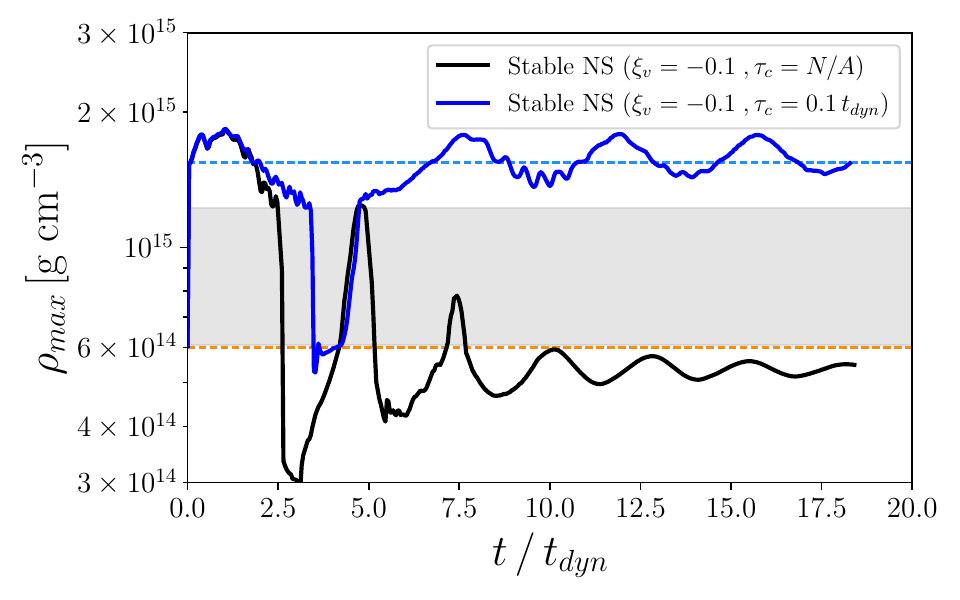}
\caption{\label{Stable_NS_to_TS}  Evolution of the maximum density of a stable NS with $M_0=1.44\,M_\odot$ and a strong initial velocity perturbation without cooling (solid black) and with a rapid cooling prescription (solid blue). Similar to Fig.~\ref{UTS_44}, the dashed lines represent the maximum density of the stable NS and stable TS with the same mass, and the shaded grey area shows the phase transition range of densities.} 
\end{figure}

Our simulations demonstrate this behavior directly.
Fig.~\ref{Stable_NS_to_TS} shows the evolution of the maximum rest-mass density for two initially identical stable NSs subjected to an inward radial velocity perturbation of $\xi_v=-0.1$. The simulations
differ only in the treatment of cooling. Without cooling, the initial compression drives the star through the phase transition region, but
the high density state is not maintained. The star subsequently re-expands and, after several dynamical timescales, returns to the NS branch. Its late time density oscillates around a value slightly below
the initial equilibrium density, reflecting the additional energy retained after the perturbation. In contrast, when rapid cooling is active throughout the evolution, the star remains on the high density side of the phase transition and ultimately oscillates around the stable TS configuration with the same rest mass. The removal of thermal energy during the transition therefore allows the initially stable NS to settle on the more gravitationally bound TS branch.

\begin{figure*}[t]
\includegraphics[width=\linewidth]{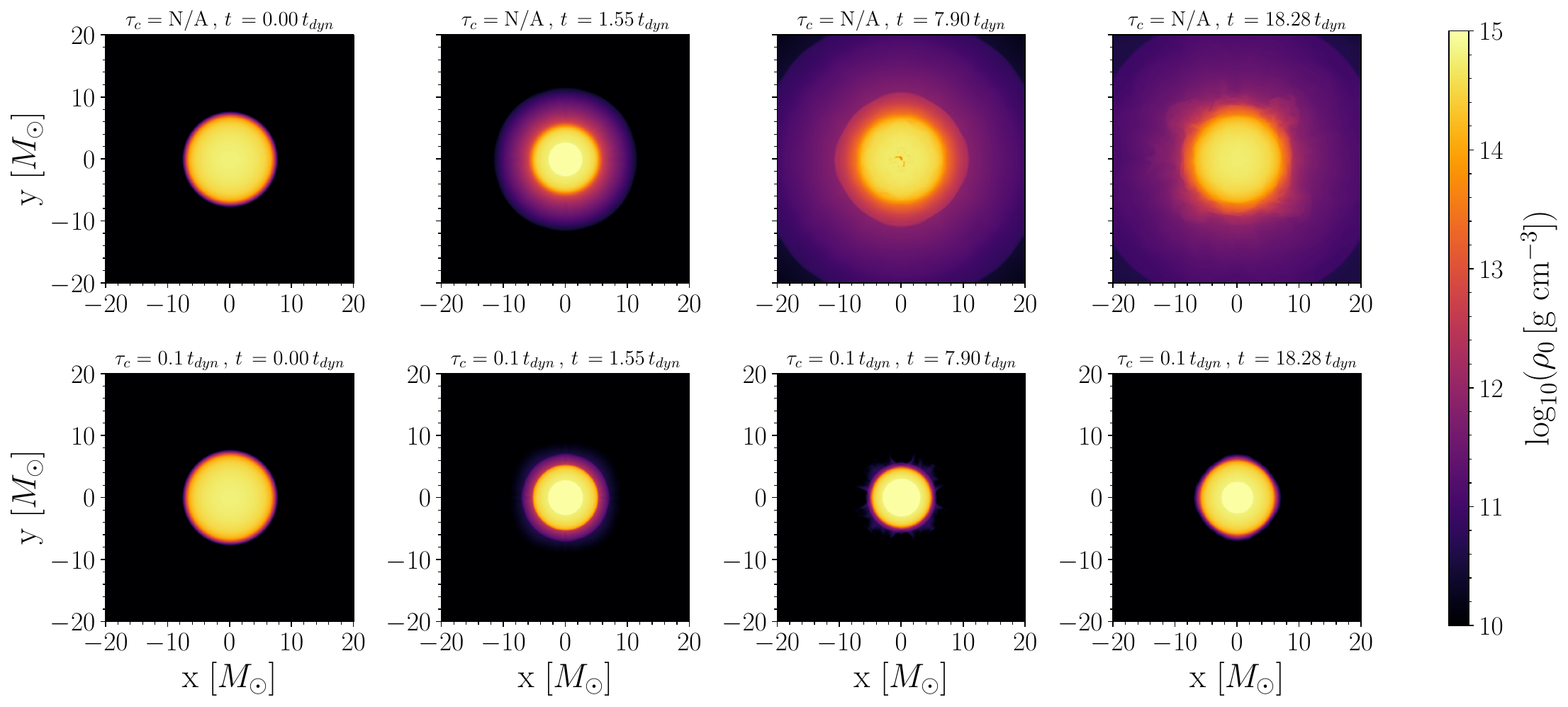}
\caption{\label{NS_cooling} Snapshots of the matter evolution of the two simulations reported in Fig.~\ref{Stable_NS_to_TS} representing a stable NS with a strong inwards velocity perturbation. The upper panels show the evolution without cooling. Although the stellar interior temporarily enters the quark matter phase (the bright region in the second plot), the generated thermal pressure drives the star back toward the NS branch. The final configuration differs from the initial cold NS because thermal energy remains in the fluid. The lower panels show the corresponding evolution with cooling, for which a quark core forms and the star settles near the stable TS configuration.} 
\end{figure*}

\begin{figure*}[t]
\includegraphics[width=\linewidth]{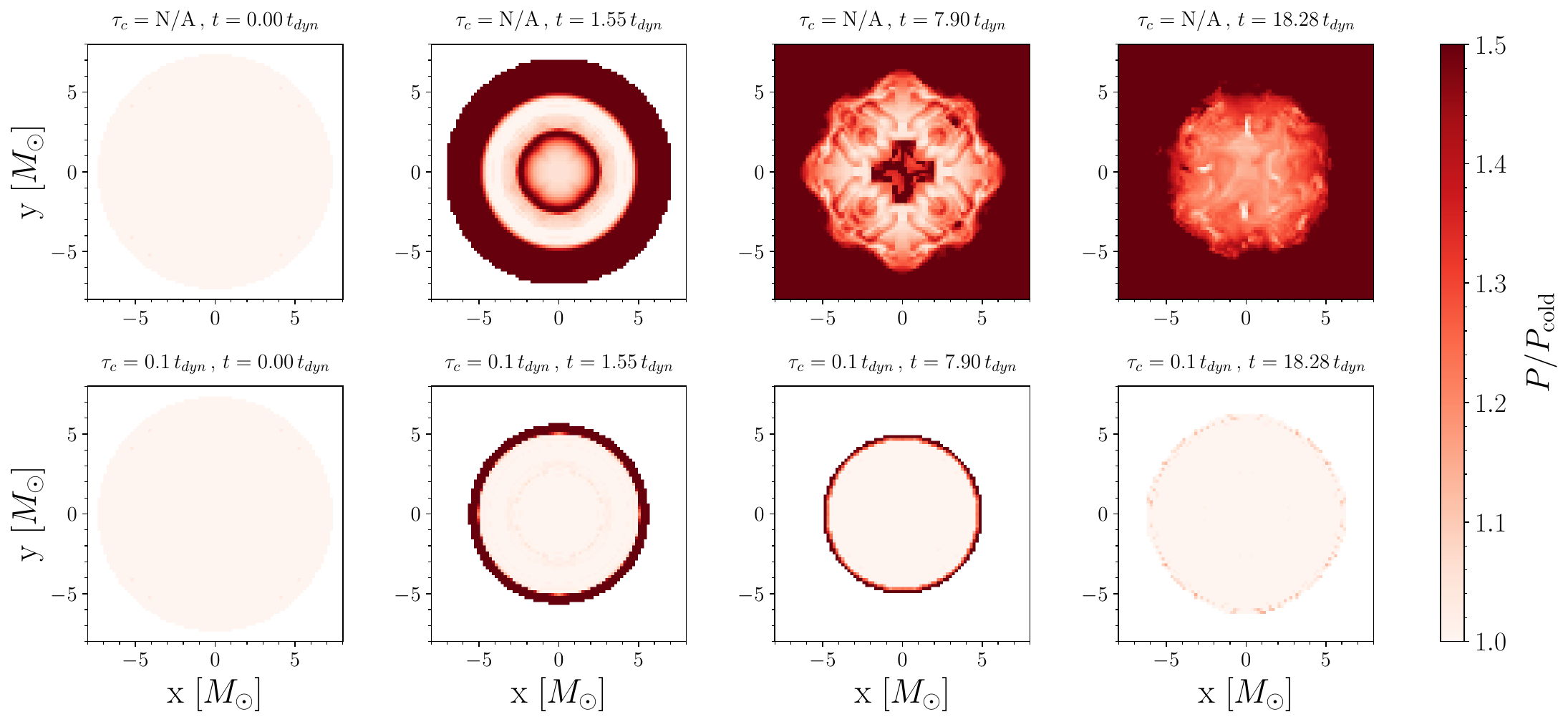}
\caption{\label{Stable_NS_cooling} Thermal evolution of the simulations shown in Fig.~\ref{NS_cooling}. In the absence of cooling, a substantial thermal pressure component develops inside the star (upper panels) due to shocks generated by the bounce, whereas cooling with $\tau_c=0.1\,t_{dyn}$ efficiently removes the generated thermal energy (lower panels).  The plot shows $P/P_{\rm cold}$ for rest-mass densities above $10^{-3}\rho_{max}$ where bulk of the stellar rest mass lies.} 
\end{figure*}

\begin{figure*}[t]
\includegraphics[width=\linewidth]{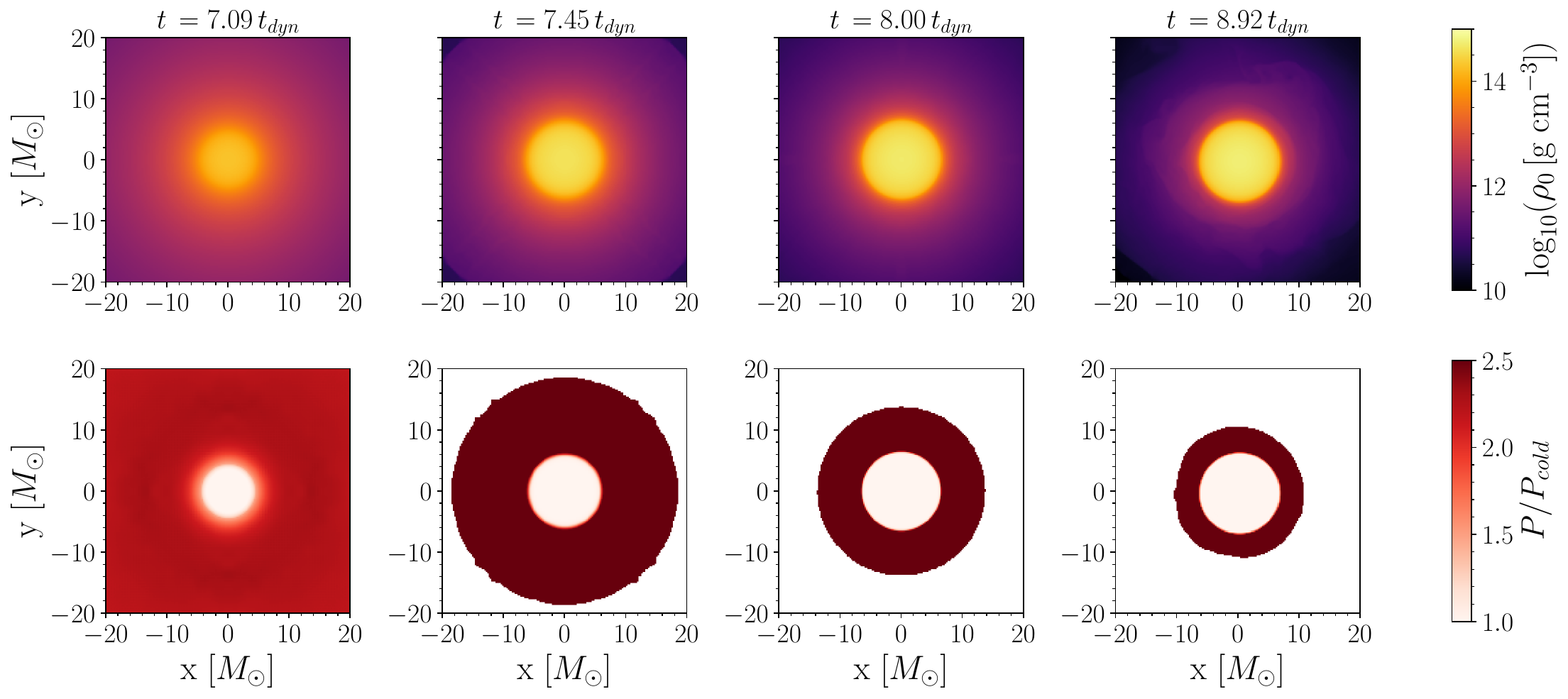}
\caption{\label{WD_cooling} Two-dimensional snapshots of the final stages of the collapse of an unstable WD with $M_0=1.44\,M_\odot$. The top plots show the matter distribution, and the bottom plots quantify thermal pressure with $P/P_{cold}$. Similarly to the previous plots of $P/P_{cold}$, we only show this ratio for rest-mass densities greater than $10^{-3}\rho_{max}$.} 
\end{figure*}

The thermal evolution helps to elucidate the different outcomes. Figs.~\ref{NS_cooling} and~\ref{Stable_NS_cooling} show snapshots of the matter distribution and $P/P_{\rm cold}$ for the two
evolutions presented in Fig.~\ref{Stable_NS_to_TS}. In the absence of cooling, compression and shocks generate a substantial thermal component as the star contracts. The resulting thermal pressure provides additional support, causing the star to re-expand and
preventing it from settling on the stable TS branch. When cooling is active, the dynamically generated thermal energy is removed while the star contracts, reducing this additional pressure support and allowing the configuration to remain at high density and relax toward the stable TS.

The adiabatic EOS adopted in
Ref.~\cite{Haque:2026ero} suppresses this thermal support by construction. The transitions in that study therefore probe the limiting case in which  thermal energy is radiated away instantaneously, whereas our simulations explicitly allow for the generation of heat and its removal is controlled through the cooling timescale. The results in Ref.~\cite{Haque:2026ero} are compatible with our extreme cooling simulations.

The two directions of migration are therefore dynamically asymmetric. A TS-to-NS transition moves the system toward a less gravitationally bound equilibrium configuration and can be facilitated by the energy supplied through the perturbation. In contrast, an NS-to-TS transition terminates in a more gravitationally bound configuration and requires the dynamically generated thermal energy to be removed sufficiently rapidly. An inward velocity perturbation can initiate the contraction and carry
the star across the phase transition region, but in our simulations, sufficiently efficient cooling is required for the resulting configuration to remain on the TS branch.

\subsubsection{Collapse of unstable white dwarfs}
The final scenario we consider is the gravitational collapse of an unstable WD. We focus on the simulation represented by the dashed black curve in Fig.~5 of Ref.~\cite{Naseri:2024rby}, with rest mass
$M_0=1.44\,M_\odot$. As shown in Fig.~\ref{E_g}, at this fixed rest mass the unstable WD is the least gravitationally bound among the  configurations with the same rest mass, while progressively more bound
configurations are encountered on the stable NS, unstable TS, and stable TS branches. The collapse therefore provides a useful test of how far the system can progress through this gravitational binding energy hierarchy as the thermal energy generated during the evolution is removed.

Fig.~\ref{WD_cooling} shows the matter and thermal evolution during the final stages of the collapse. We focus on the late time evolution at $t\gtrsim7.09\,t_{\rm dyn}$, when cooling is activated in the simulation. Prior to this time, the collapse generates a substantial thermal component through compression. The resulting
thermal pressure halts the contraction and produces an extended pressure-supported region surrounding the dense core, visible as the bright halo in $P/P_{\rm cold}$ in Fig.~\ref{WD_cooling}. Once cooling
is activated, this thermal support is progressively removed and the collapse resumes, allowing the cold, high-density core to grow.

At late times, the excess thermal component has been almost entirely removed and the remnant approaches the cold stable NS configuration with the same rest mass. Importantly, reaching the cold NS also exhausts the energy reservoir accessible to the cooling prescription; as $e_{\rm th}\rightarrow0$, the emissivity $\Lambda=\rho_0e_{\rm th}/\tau_c$ vanishes. The adopted cooling mechanism can therefore no longer extract additional energy from the star or drive further contraction toward the more gravitationally bound stable TS. Within this model, reaching the TS would require the system to lose more energy than is contained in the dynamically
generated thermal component. Continuing the phenomenological cooling prescription beyond this point would require reducing the internal energy below its zero-temperature EOS value and would therefore no longer represent physical thermal cooling.

This provides a physical explanation for why the collapse simulations of Ref.~\cite{Naseri:2024rby} robustly produce stable NSs despite the existence, at the same rest mass, of a more gravitationally bound stable TS. Cooling allows the collapsing WD to shed the thermal energy generated during the collapse and approach the cold equilibrium sequence, but the removable thermal energy reservoir is exhausted once the remnant reaches the stable NS. The existence of a more gravitationally bound equilibrium configuration alone is therefore not sufficient to guarantee that it is dynamically accessible through this formation channel.

\subsection{Discussion and astrophysical implications}
The numerical experiments presented in this paper provide a unified energetic interpretation of the different dynamical outcomes reported in
Refs.~\cite{Espino:2021adh,Naseri:2024rby,Haque:2026ero}. The central point is that the existence of several equilibrium configurations at the same rest mass does not imply that all of them are equally
accessible dynamically. As shown in Fig.~\ref{E_g}, at fixed $M_0$ the gravitational binding energy systematically decreases from the unstable WD to the stable NS, unstable TS, and stable TS configurations. The stable TS is therefore the most gravitationally bound equilibrium configuration among these possibilities. However, reaching this state through the dynamical pathways considered in this work requires not only driving the star toward higher density, but also preventing the thermal
energy generated during the transition from halting or reversing the contraction.

This distinction is particularly important because dynamical transitions do not proceed along the cold TOV sequence and are not adiabatic. Contraction, compression, phase transitions, and shocks convert part of the available energy into internal energy and generate thermal pressure. In the absence of cooling, this energy remains in the fluid and can provide sufficient support to reverse the contraction, even after the star has entered or
crossed the phase transition region. Consequently, the presence of a more gravitationally bound cold
equilibrium configuration with the same rest mass  does not by itself guarantee that the system will settle there. Its dynamical accessibility depends on whether the
thermal energy generated during the transition can be removed before the associated pressure support reverses the contraction.

The results of the unstable TS simulations make this competition particularly clear. Starting from the unstable branch, the star can initially evolve toward the more gravitationally bound TS configuration. Whether it remains there depends on how rapidly the thermal energy generated during this contraction is removed. When cooling operates sufficiently rapidly, the star can remain at high density and relax toward the stable TS. For slower cooling, thermal pressure causes re-expansion before this relaxation occurs, and the star instead migrates toward the stable NS. The required cooling becomes increasingly rapid toward configurations with lower $|E_t|$, reaching $\tau_c\lesssim t_{\rm dyn}$, and in some cases substantially below $t_{\rm dyn}$, in our simulations.

The transitions between initially stable NSs and TSs further illustrate the asymmetry implied by this picture. A transition from a stable TS to a stable NS moves the system toward a less gravitationally bound equilibrium configuration, and the energy
injected by a sufficiently strong perturbation can facilitate such a transition. The reverse transition is dynamically different. An inward velocity perturbation can compress a stable NS and drive it
across the phase transition region, but it simultaneously adds energy to the system, part of which is converted into thermal energy during the subsequent evolution. For the star to remain on the more gravitationally bound TS branch, this dynamically generated thermal energy must be removed. This explains why including the dynamically generated thermal component changes the outcome relative to the cold pressure experiments of Ref.~\cite{Haque:2026ero}, and why sufficiently rapid cooling allows the NS-to-TS transition to occur again in our simulations.

The collapse of an unstable WD provides the complementary limiting case. Here, cooling is necessary for the collapse to proceed after thermal pressure halts the initial contraction. Nevertheless, cooling does not drive the system indefinitely toward configurations of increasing gravitational binding energy. Once the dynamically generated
thermal energy has been radiated away and the remnant reaches the cold NS branch, $e_{\rm th}\rightarrow0$ and the cooling emissivity vanishes. No further thermal energy is available to remove, and the system cannot continue toward the still more gravitationally bound TS through the same mechanism. Thus, the final state is determined not only by the ordering of equilibrium configurations in $E_g$, but also by the amount of removable energy generated during the dynamical transition.

These results suggest that the dynamical accessibility of the TS branch is controlled by the competition between the cooling and dynamical timescales. The thermal energy generated during the
transition is removed on a characteristic timescale $\tau_c$, whereas the contraction and subsequent re-expansion occur on the dynamical timescale. For cooling to substantially alter the outcome of a rapid migration, it must remove the generated thermal support before the configuration re-expands. Our simulations therefore suggest the approximate condition $\tau_c \lesssim t_{\rm dyn}$, with substantially shorter cooling timescales required for some of the lower-mass configurations considered here.

It is insightful to compare this requirement with an estimate of the physical cooling timescale associated with the heating produced during the migration. In stable configurations, the central value of $e_{\rm th}$ remains small throughout the evolution, indicating that no appreciable physical heating occurs. The unstable TS, in contrast, develops a substantial thermal
component during its migration. For example, in the case shown by the solid black curve in Fig.~\ref{UTS_44}, the central thermal specific  energy increases rapidly from $e_{\rm th}\sim10^{-6}$ to $e_{\rm th}\sim10^{-2}$ at $t\approx9.5\,t_{\rm dyn}$. The dramatic increase during the unstable evolution therefore represents genuine heating associated with the dynamical migration.

To obtain an order-of-magnitude estimate of the corresponding temperature, we follow Ref.~\cite{Paschalidis:2012ff} and use
\begin{equation}
\label{temperature}
e_{\rm th}
 = \frac{3k_B T}{2m}
 + f\frac{aT^4}{\rho_0},
\end{equation}
where $k_B$ and $a$ are the Boltzmann and radiation constants, respectively. In the hadronic phase we take $m=m_n$, while after deconfinement we adopt
$m=m_n/3$ as an effective energy scale per quark ($m_n$ stands for the neutron mass). The factor $f$ parametrizes the relativistic degrees of freedom, with $f=1$ when photons dominate and $f=43/8$ at high temperatures to account approximately for electron--positron pairs and trapped neutrinos and antineutrinos~\cite{Paschalidis:2011ez}. Equation~\eqref{temperature} is intended only as an approximate temperature diagnostic. In particular,
degenerate quark and hadronic matter is not accurately described by the classical ideal gas term~\cite{Raithel:2019gws,Raithel:2021hye}, and a quantitative temperature determination would require a finite temperature hadron--quark EOS and the corresponding composition.

Using this temperature, the neutrino diffusion timescale can be estimated as~\cite{Paschalidis:2012ff},
\begin{equation}
\label{cooling_estimate}
\tau_\nu \approx
400
\left(\frac{M}{1.4\,M_\odot}\right)
\left(\frac{R}{10\,{\rm km}}\right)^{-1}
\left(\frac{E_\nu}{10\,{\rm MeV}}\right)^2
\,{\rm ms},
\end{equation}
where $E_\nu$ is the characteristic neutrino energy, and $R$ is the stellar radius. Taking $E_\nu\approx3.15\,k_BT$ for a thermal neutrino
distribution~\cite{Nakazato:2018xkv} and applying
Eqs.~\eqref{temperature} and \eqref{cooling_estimate} to the heated
configuration gives $\tau_\nu\sim10^4\,{\rm ms}$. Although this estimate is necessarily approximate, the resulting timescale is many orders of magnitude longer than the stellar dynamical timescale (e.g.~$t_{dyn}=0.17\,\textrm{ms}$, for this star).

This large separation of timescales is consistent with detailed studies of NS thermal evolution, in which neutrino and photon emission and thermal
transport operate on timescales far longer than the dynamical timescale~\cite{Yakovlev:2000jp,Yakovlev:2004iq}.
The thermal energy generated during the migration is therefore expected to remain trapped over the few dynamical timescales that determine whether the
star remains at high density or re-expands. Within the formation channels investigated here, this timescale hierarchy favors the NS branch; although a
stable TS is more gravitationally bound at the same rest mass, reaching it generally requires thermal energy to be removed much more rapidly than
expected from realistic cooling processes. Nevertheless, formation of TSs through other astrophysical channels, including mass loss in HSs with masses slightly larger than the TS range~\cite{Naseri:2024rby}, could still be promising.

\section{\label{conclusions}Conclusions} 
In this work, we investigated the role of gravitational binding energy in the dynamical accessibility of equilibrium configurations when a strong first-order phase transition produces a separate TS branch. At fixed rest mass, we find that the gravitational binding energy becomes
increasingly negative toward the more compact configurations, with the stable TS being more gravitationally bound than the corresponding stable
NS. However, our simulations demonstrate that this energetic ordering alone does not determine the dynamical outcome. During rapid transitions,
contraction and shocks generate thermal energy and pressure that can halt or reverse the evolution toward the more compact TS branch.

General relativistic hydrodynamical simulations of unstable TS migration, transitions between stable NSs and TSs, and unstable WD collapse reveal a
common picture: access to the more gravitationally bound TS branch depends on whether the thermal energy generated during the transition can be removed
before the star re-expands. Sufficiently rapid cooling can allow the star to remain at high density and settle onto the stable TS branch, whereas slower
or absent cooling generally favors an NS remnant. For the configurations studied here, TS formation through these rapid dynamical pathways generally
requires cooling on a timescale comparable to or shorter than the stellar dynamical timescale, $\tau_c\lesssim t_{\rm dyn}$, with even shorter
timescales required for some lower-mass configurations.

Realistic cooling mechanisms in NSs operate on timescales much longer than $t_{\rm dyn}$; hence, the thermal energy generated during a
rapid transition is expected to remain largely trapped over the timescale that determines the immediate dynamical outcome. Within the constant-rest-mass formation
channels investigated here this timescale hierarchy therefore favors the NS branch even when a more gravitationally bound stable TS exists at the same rest mass. This does not preclude TS formation through other astrophysical channels with different thermal or dynamical conditions or ones that do not conserve rest-mass, but
it demonstrates that the existence of a stable TS equilibrium configuration that is more gravitationally bound does not by itself guarantee its dynamical accessibility.

A few caveats are in order. First, the results in this work apply to a single EOS with a hadron-quark phase transition. Moreover, we adopt \(\Gamma_{\rm th}=2\). At fixed thermal energy, \(\Gamma_{\rm th}-1\) controls the efficiency with which dynamically generated thermal energy contributes to pressure support. A smaller value of \(\Gamma_{\rm th}\) would reduce this support and could quantitatively shift the critical cooling timescale or the outcome of cases close to the transition boundary. Nevertheless,  even  \(\Gamma_{\rm th}=1.75\) for the hadronic matter and \(\Gamma_{\rm th}=1.333\) for the quark matter would still generate significant heat~\cite{Blacker:2024tet}. Therefore, the qualitative conclusion—that retained thermal energy can prevent access to cold twin stars—should remain unchanged. Establishing quantitative transition thresholds requires a finite-temperature hadron–quark EOS and lies beyond the scope of this work, whose aim is to make a point of principle and to shed light on prior simulations. Our simulations support the point of principle we make -- the existence of equilibrium configurations that are more gravitationally bound does not by itself guarantee their dynamical accessibility. This situation is analogous to the existence of WDs in the NS mass range; WDs can be the dynamically realized outcome for certain progenitors even though more gravitationally bound NS configurations exist at the same rest mass. Second, our EOS does not have a temperature-dependent hadron-to-quark phase transition density. Depending on the details of the nuclear physics, this could impact both the dynamical evolution and the determination that no quarks exist in the configurations we identify as oscillating around the NS branch in the TS mass range. This is because in the QCD phase diagram it is known that the quark deconfinement density decreases with increasing temperature. However, for some finite-temperature hybrid EOSs, the density threshold for quark deconfinement depends only mildly on temperature. For example, for $ T\leq 20$ MeV (as in our simulations) the density threshold for quark deconfinement in the EOS of~\cite{Bauswein:2018bma} changes by less than $\simeq 10\%$. Therefore, we expect that our conclusions apply for at least some realistic nuclear EOSs with a strong hadron-quark phase transition and that for these EOSs the conclusion that a configuration oscillates around the corresponding stable NS remains valid. Third, our temperature estimates are crude but give the correct order of magnitude, and it is well known that NS cooling timescales are on the order of $\mathcal{O}(1s)$. Finally, our adopted effective cooling is crude, but it is precisely the tool we need to make our point of principle since we can control the cooling timescale. More realistic neutrino cooling and realistic finite-temperature EOSs are important for quantitative conclusions, but we expect that none of these will change our conclusions qualitatively.

\begin{acknowledgments}
We thank Thomas Baumgarte for useful discussions while this work was in development, and for suggesting that the binding energy may be one way to distinguish whether neutron stars or twin stars are the preferred dynamical outcome. This work was in part supported by NASA grants 80NSSC24K0771, 80NSSC26K0343, and NSF grant PHY-2145421 to the University of Arizona. The numerical simulations were performed in part using the Puma and Ocelote high-performance computing systems at the University of Arizona. The work was also supported by allocation PHY190020 from the Advanced Cyberinfrastructure Coordination Ecosystem: Services \& Support (ACCESS) program, which is supported by U.S. National Science Foundation grants 2138259, 2138286, 2138307, 2137603, and 2138296.
\end{acknowledgments}

\bibliography{apssamp}

\end{document}